\documentclass[12pt]{article}
\pdfoutput=1
\usepackage{amsmath}
\usepackage{graphicx}
\usepackage{caption}
\usepackage{subcaption}
\usepackage{tabularx}
\usepackage{bigstrut}
\usepackage{multirow}
\usepackage[utf8]{inputenc}
\usepackage[pdfauthor={Declan Millar}, pdftitle={Using asymmetry observables to discover and distinguish Z' signals in top pair production with the lepton-plus-jets final state at the LHC}, pdfkeywords={Z' zprime boson top quark pair production charge spin asymmetry asymmetries lepton-plus-jets},hidelinks]{hyperref}
\usepackage[numbers,compress]{natbib}
\usepackage{float}

\def\Title#1{\begin{center} {\Large #1 } \end{center}}
\def\Author#1{\begin{center}{ \sc #1} \end{center}}
\def\Address#1{\begin{center}{ \it #1} \end{center}}

\newcommand\pubblock{\rightline{\begin{tabular}{l} \pubnumber\\
         \pubdate  \end{tabular}}}
\newenvironment{Abstract}{\begin{quotation}  }{\end{quotation}}
\newenvironment{Presented}{\begin{quotation} \begin{center} 
             PRESENTED AT\end{center}\bigskip 
      \begin{center}\begin{large}}{\end{large}\end{center} \end{quotation}}
\def\Acknowledgements{\bigskip  \bigskip \begin{center} \begin{large}
             \bf ACKNOWLEDGEMENTS \end{large}\end{center}}

\def\beq{\begin{equation}}
\def\eeq#1{\label{#1}\end{equation}}
\def\eeqn{\end{equation}}

\def\beqa{\begin{eqnarray}}
\def\eeqa#1{\label{#1}\end{eqnarray}}
\def\eeqan{\end{eqnarray}}

\let\bar=\overbar

\def\Dslash{\not{\hbox{\kern-4pt $D$}}}
\def\dslash{\not{\hbox{\kern-2pt $\del$}}}

\def\msb{{\bar{\ssstyle M \kern -1pt S}}}

\newcommand\pubnumber{SNSN-323-63}
\newcommand\pubdate{November 30, 2016}

\begin{document}

\pagenumbering{Alph}
 
\begin{titlepage}
  \pubblock
  \vfill
  \Title{Discriminating \texorpdfstring{$Z'$}{Z'} signals in semileptonic top pair production at the LHC}
  \vfill
  \Author{Lucio Cerrito$^1$, Declan Millar$^{2,3}$,\\ Stefano Moretti$^3$, Francesco Span\`{o}$^4$}
  \Address{$^1$Department of Physics, University of Rome Tor Vergata and INFN, Via Della Ricerca Scientifica, 1, Rome, 00133, Italy}
  \Address{$^2$School of Physics and Astronomy, Queen Mary University of London, Mile End Road, London E1 4NS, United Kingdom}
  \Address{$^3$School of Physics and Astronomy, University of Southampton, Highfield, Southampton SO17 1BJ, United Kingdom}
  \Address{$^4$Department of Physics, Royal Holloway University of London, Egham Hill, Egham TW20 0EX, United Kingdom}
  \begin{Abstract}
    We investigate the sensitivity of top pair production to the properties of different Beyond the Standard Model theories embedding a new neutral boson. We include six-fermion decay, and account for the full tree-level Standard Model $t\bar{t}$ interference, with all intermediate particles allowed off-shell. We focus on those observables best suited to the lepton-plus-jets final state at the LHC, and simulate the resulting experimental conditions, including kinematic requirements and top quark pair reconstruction in the presence of missing transverse energy and combinatorial ambiguity in quark-top assignment. In particular, we demonstrate the use of asymmetry observables to probe the coupling structure of a new neutral resonance, in addition to cases in which these asymmetries may even form complementary discovery observables in combination with the differential cross section.
  \end{Abstract}
  \vfill
  \begin{Presented}
    $9^{th}$ International Workshop on Top Quark Physics\\
    Olomouc, Czech Republic,  September 19--23, 2016
  \end{Presented}
  \vfill

  \thispagestyle{empty}

\end{titlepage}

\pagenumbering{arabic}

\section{Introduction}
\label{sec:introduction}

New fundamental, massive, neutral, spin-1 gauge bosons ($Z'$) appear ubiquitously in theories Beyond the Standard Model (BSM). Resonance searches in the $t\bar{t}$ channel can offer unique handles on the properties of a $Z'$, due to asymmetry observables available because (anti)tops decay prior to hadronisation and their spin information is cleanly transmitted to their decay products. Their definition in $t\bar{t}$, however, requires the reconstruction of the top quark pair. We simulate top pair production and six-fermion decay mediated by a $Z'$, with analysis focused on the lepton-plus-jets final state, and imitate some resulting experimental conditions at the parton level, in order to assess the prospect for an LHC analysis to discover and distinguish a $Z'$ boson using a combination of these observables~\cite{cerrito2016}.

\section{Models}
\label{sec:models}

There are several candidates for a Grand Unified Theory (GUT), a hypothetical enlarged gauge symmetry, motivated by gauge coupling unification at approximately the $10^{16}$ GeV energy scale. $Z'$ often arise due to the residual U$(1)$ gauge symmetries after their spontaneous symmetry breaking to the familiar SM gauge structure. We study a number of benchmark examples that may be classified into three types: $E_6$ inspired models, generalised Left-Right (GLR) symmetric models and General Sequential Models (GSMs)~\cite{accomando2011}. In each case two U$(1)$ symmetries survive down to the TeV scale, and for each class we may take a general linear combination of the appropriate operators, fixing $g'$ and varying the angular parameter dictating the relative strengths of the component generators, until we recover interesting limits. These models all couple universally by generation, such that the strongest experimental limits come from the DY channel. The limits for these models have been extracted based on DY results, at $\sqrt{s}=7$ and $8$~TeV with an integrated luminosity of $L=20$~fb$^{-1}$, from the CMS collaboration (with similar results from ATLAS), by Accomando et al., with such a state generally excluded below $3$ TeV~\cite{thecmscollaboration2015,theatlascollaboration2014c,accomando2016}.

\section{Method}
\label{sec:method}

Measuring $\theta$ as the angle between the top and the incoming quark direction, in the parton centre of mass frame, we define the forward-backward asymmetry:
\begin{equation}
	A_{FB}=\frac{N_{t}(\cos\theta>0)-N_t(\cos\theta<0)}{N_t(\cos\theta>0)+N_t(\cos\theta<0)}, \quad \cos\theta^* =\frac{y_{tt}}{|y_{tt}|}\cos\theta 
\end{equation}
With hadrons in the initial state, the quark direction is indeterminate. However, the $q$ is likely to carry a larger partonic momentum fraction $x$ than the $\bar{q}$ in $\bar{x}$. Therefore, to define $A^{*}_{FB}$ we choose the $z^*$ axis to lie along the boost direction. The top polarisation asymmetry ($A_{L}$), measures the net polarisation of the (anti)top quark by subtracting events with positive and negative helicities:
\begin{equation}
  A_{L}=\frac{N(+,+)+N(+,-)-N(-,-)-N(-,+)}{N(+,+)+N(+,-)+N(-,-)+N(-,+)}, \quad \frac{1}{\Gamma_l}\frac{d\Gamma_l}{dcos\theta_l}=\frac{1}{2}(1 + A_L \cos\theta_l),
\end{equation}
where $\lambda_{t}$($\lambda_{\bar{t}}$) denote the eigenvalues under the helicity operator of $t$($\bar{t}$). Information about the top spin is preserved in the distribution of $\cos\theta_l$.

In each of the models, the residual U$(1)'$ gauge symmetry is broken around the TeV scale, resulting in a massive $Z'$ boson. This leads to an additional term in the low-energy Lagrangian, from which we may calculate the unique $Z'$ coupling structure for each observable:
\begin{align}
  \mathcal{L} &\supset g^\prime Z^\prime_\mu \bar{\psi}_f\gamma^\mu(f_V - f_A\gamma_5)\psi_f,
  \label{eq:zprime_lagrangian}\\
  \hat{\sigma} &\propto \left(q_V^2 + q_A^2\right)\left((4 - \beta^2)t_V^2 + t_A^2\right),\\
  A_{FB} &\propto q_V q_A t_V t_A,\\
  A_{L} &\propto \left(q_V^2 + q_A^2\right)t_V t_A,
\end{align}
where $f_V$ and $f_A$ are the vector and axial-vector couplings of a specific fermion ($f$).

While a parton-level analysis, we incorporate restraints encountered with reconstructed data, to assess, in a preliminary way, whether these observables will survive. The collider signature for our process is a single $e$ or $\mu$ produced with at least four jets, in addition to missing transverse energy ($E^{\rm miss}_{T}$). Experimentally, the $b$-tagged jet charge is indeterminate and there is ambiguity in $b$-jet (anti)top assignment. We solely identify $E^{\rm miss}_{T}$ with the transverse neutrino momentum. Assuming an on-shell $W^\pm$ we may find approximate solutions for the longitudinal component of the neutrino momentum as the roots of a quadratic equation. In order to reconstruct the event, we account for bottom-top assignment and $p_z^\nu$ solution selection simultaneously, using a chi-square-like test, by minimising the variable $\chi^2$:
\begin{equation}
  \chi^2 = \left(\frac{m_{bl\nu}-m_{t}}{\Gamma_t}\right)^2 + \left(\frac{m_{bqq}-m_{t}}{\Gamma_t}\right)^2,
  \label{eq:chi2}
\end{equation}
where $m_{bl\nu}$ and $m_{bqq}$ are the invariant mass of the leptonic and hadronic (anti)top, respectively.

In order to characterise the sensitivity to each of these $Z'$ models, we test the null hypothesis, which includes only the known $t\bar{t}$ processes of the SM, assuming the alternative hypothesis ($H$), which includes the SM processes with the addition of a single $Z'$, using the profile Likelihood ratio as a test statistic, approximated using the large sample limit, as described in~\cite{cowan2011}. This method is fully general for any $n$D histogram, and we test both $1$D histograms in $m_{tt}$, and $2$D in $m_{tt}$ and the defining variable of each asymmetry to assess their combined significance.

\section{Results \& Discussion}
\label{sec:results}

\begin{figure}
  \centering
  \begin{subfigure}{0.494\textwidth}
    \includegraphics[width=\textwidth]{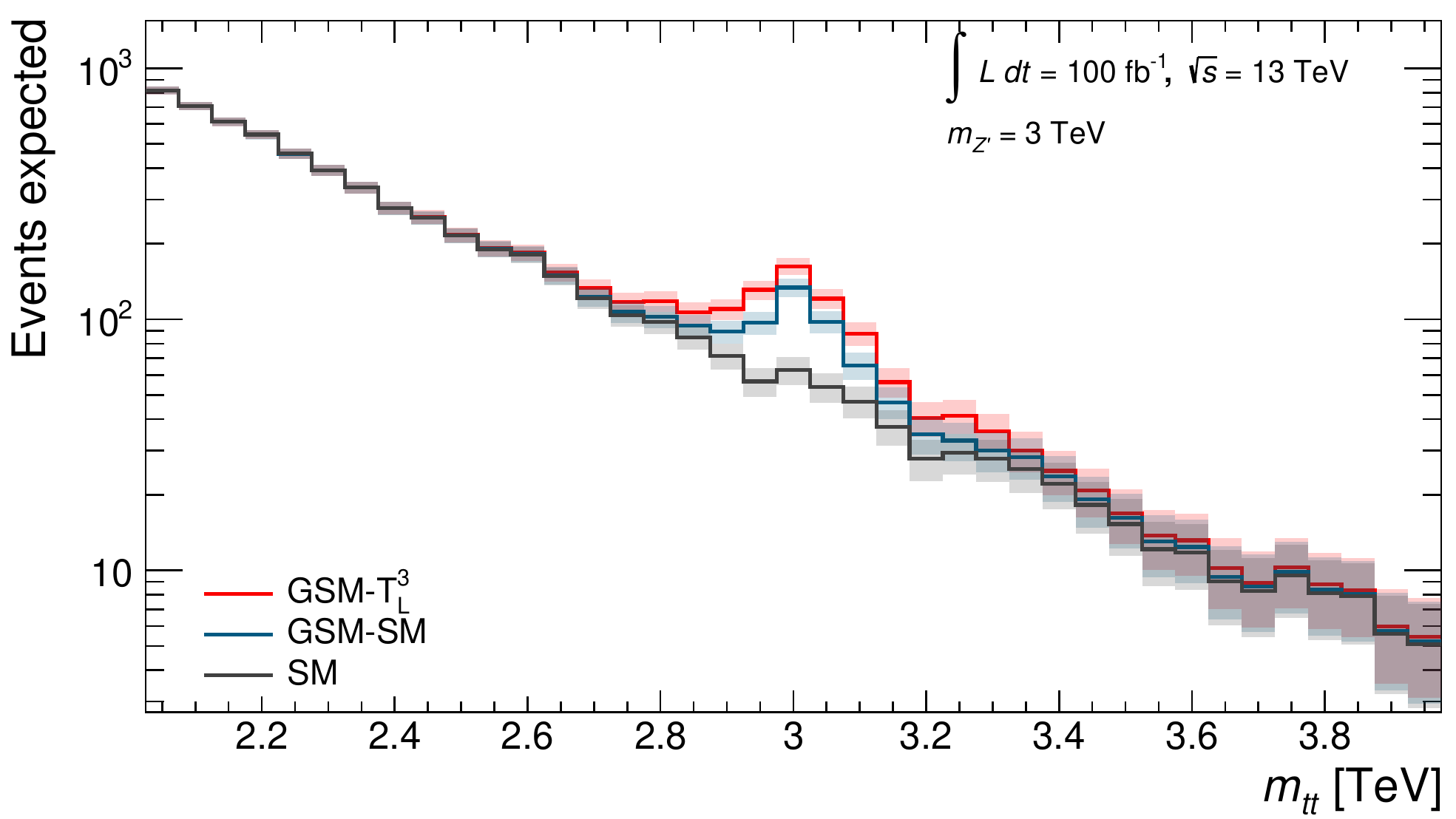}
    \caption{Events expected - GSM models}
  \end{subfigure}
  \begin{subfigure}{0.494\textwidth}
    \includegraphics[width=\textwidth]{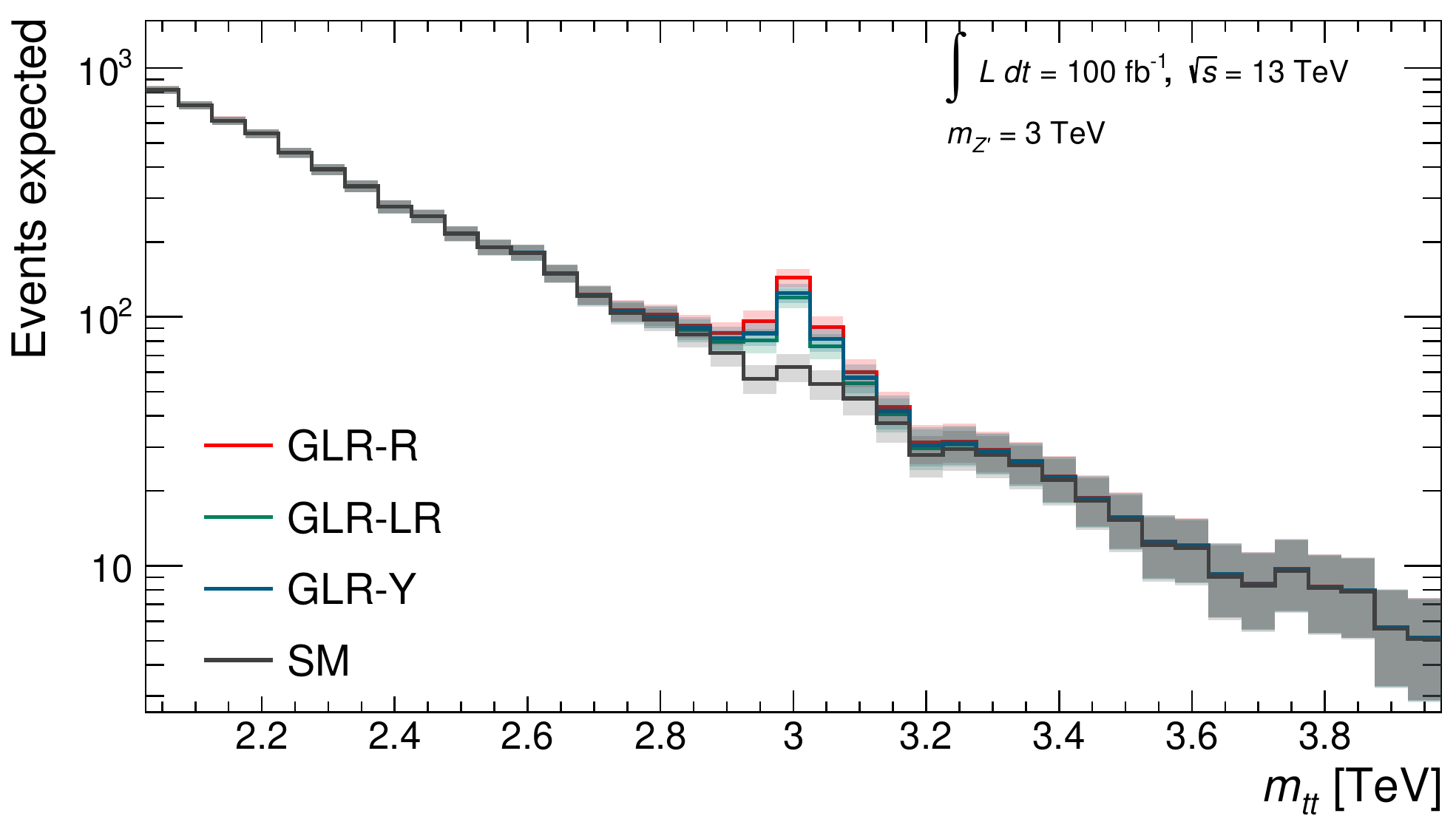}
    \caption{Events expected - GLR models}
  \end{subfigure}
  \begin{subfigure}{0.494\textwidth}
    \includegraphics[width=\textwidth]{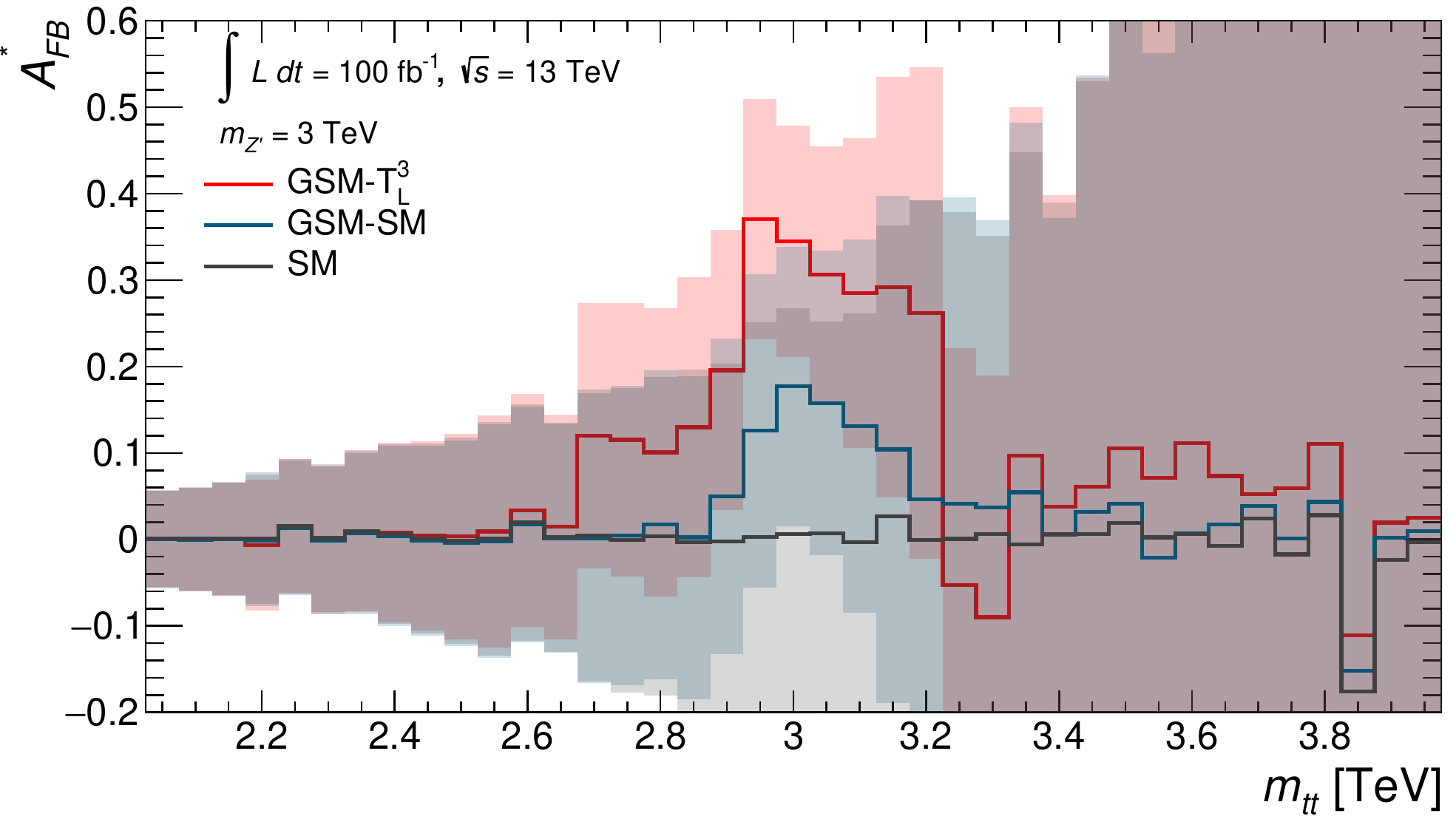}
    \caption{$A_{FB}^{*}$ - GSM models}
  \end{subfigure}
  \begin{subfigure}{0.494\textwidth}
    \includegraphics[width=\textwidth]{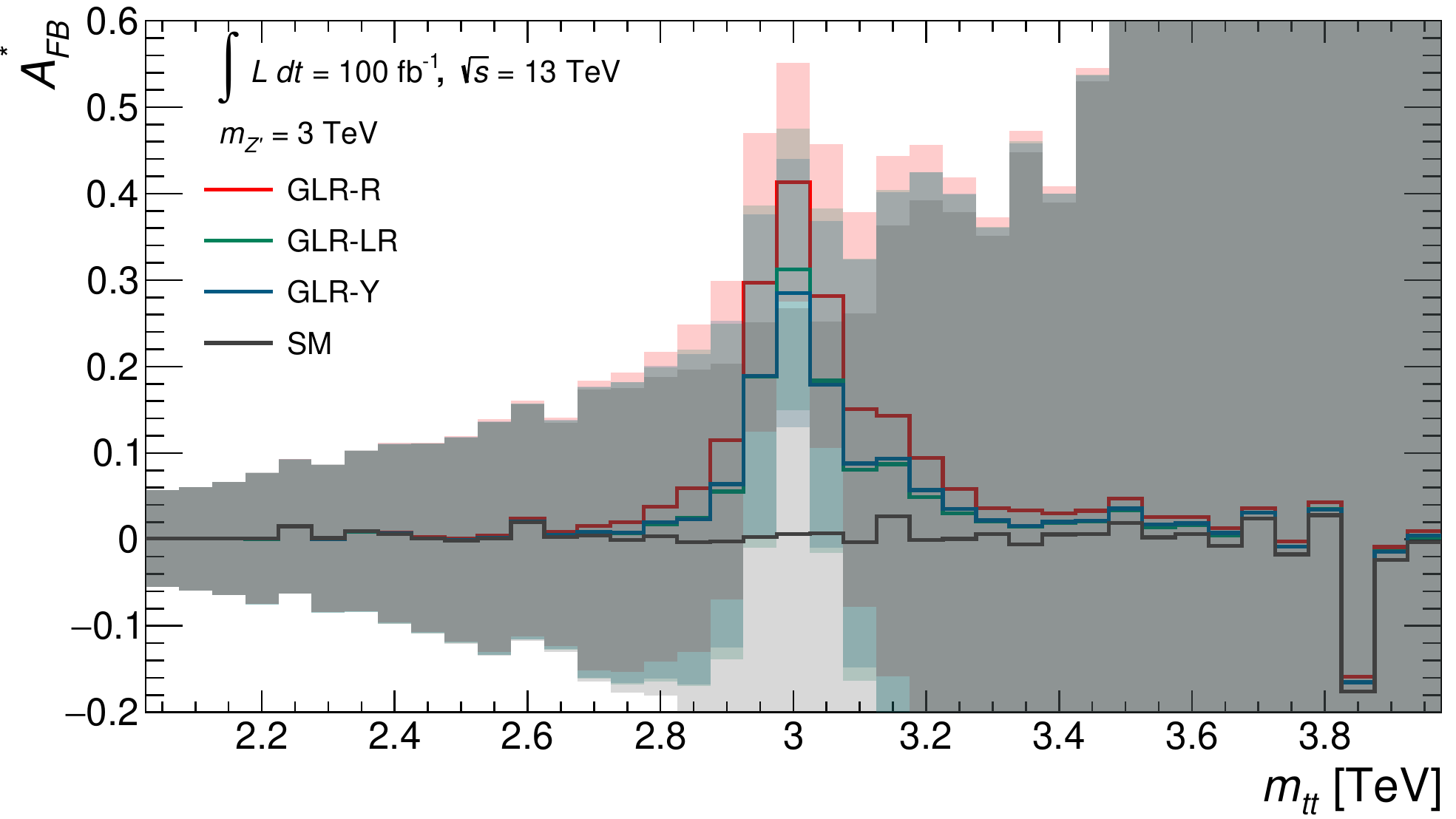}
    \caption{$A_{FB}^{*}$ - GLR models}
  \end{subfigure}
  \begin{subfigure}{0.494\textwidth}
    \includegraphics[width=\textwidth]{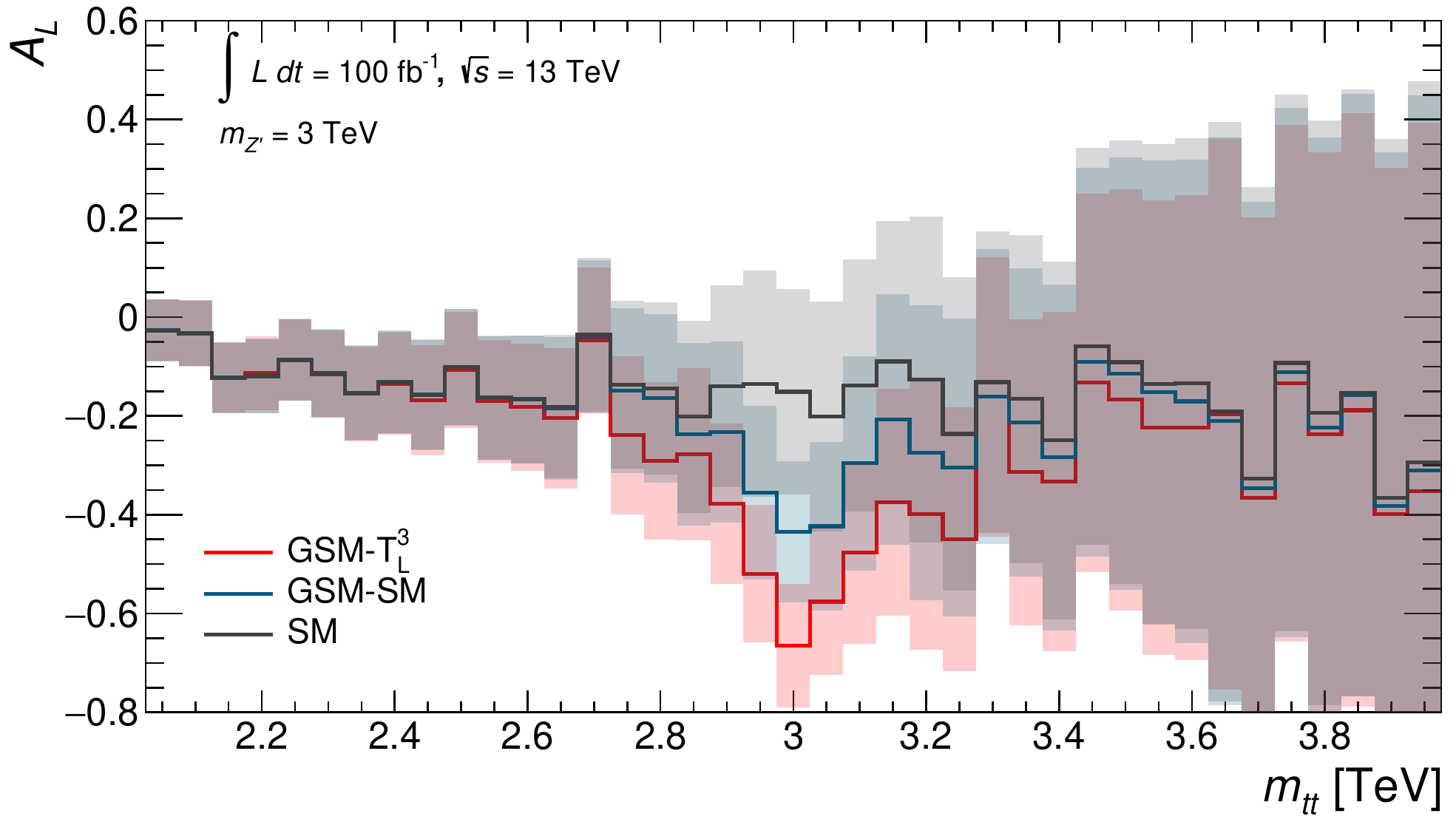}
    \caption{$A_L$ - GSM models}
  \end{subfigure}
  \begin{subfigure}{0.494\textwidth}
    \includegraphics[width=\textwidth]{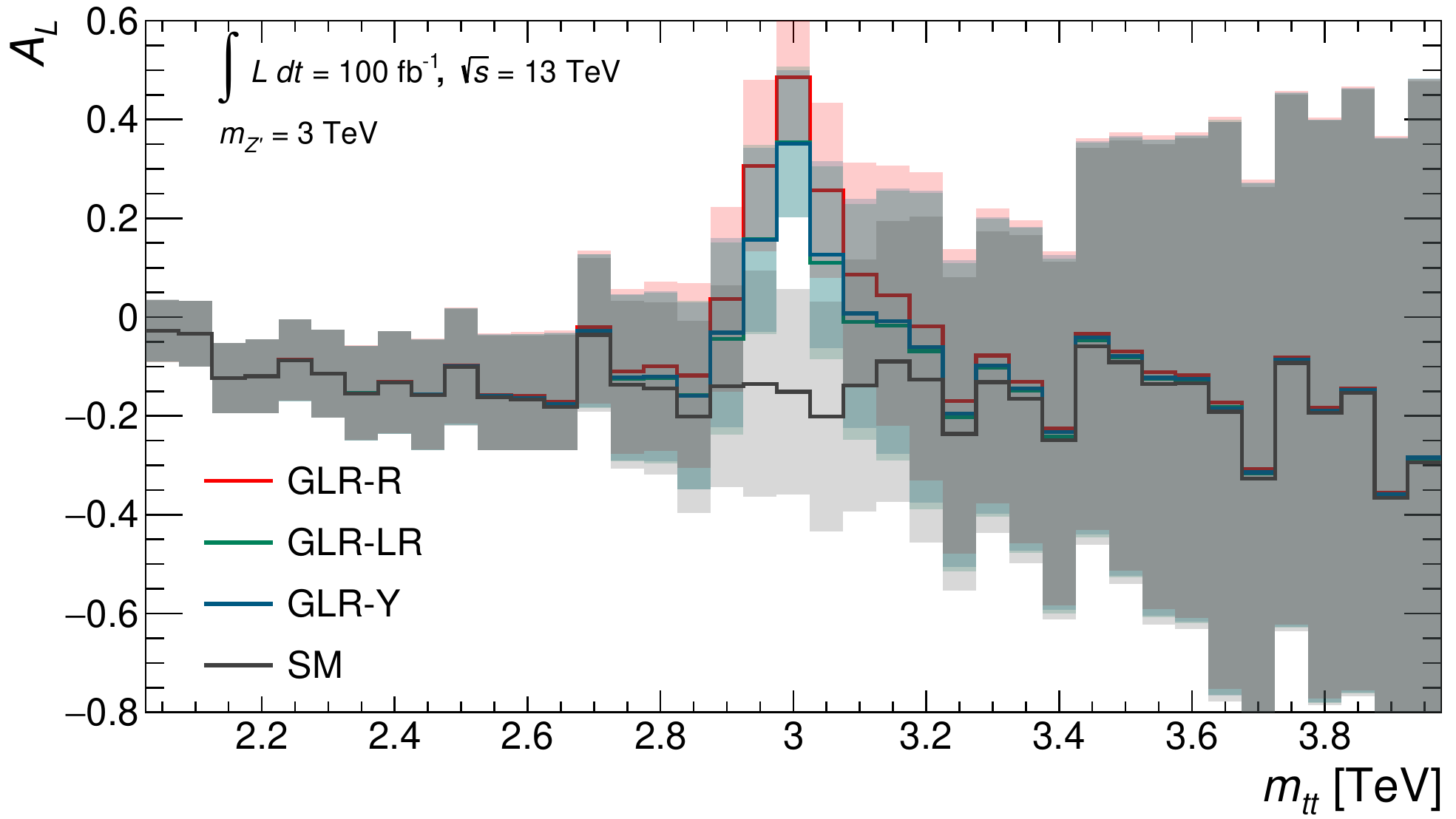}
    \caption{$A_L$ - GLR models}
  \end{subfigure}
  \caption{Expected distributions for each of our observables of interest, with an integrated luminosity of $100$~fb$^{-1}$, at $\sqrt{s}=13$~TeV. The shaded bands indicate the projected statistical uncertainty for this luminosity assuming Poisson errors. The cross section, profiled in $m_{tt}$, while having a clearly visible peak for all models, has a similar impact for all classes. Mirroring the cross section, the $A_{FB}^{*}$ distribution clearly distinguishes between the models and SM, however, the best distinguishing power of all the models investigated comes from the $A_{L}$ distribution, which features an oppositely signed peak for the GLR and GSM classes.}
  \label{fig:distinguishing}
\end{figure}

Figure~\ref{fig:distinguishing} shows plots for the differential cross section, $A_{FB}^{*}$ and $A_{L}$. The absent models, including all of the $E_6$ class have an undetectable enhancement with respect to the SM yield, and the absence of a corresponding peak in either asymmetry offers an additional handle on diagnosing a discovered $Z'$. To evaluate the usefulness of each asymmetry as a combined discovery observable we bin in both $m_{tt}$ and its defining variable. The final results of the likelihood-based test, as applied to each model, and tested against the SM, are presented in table~\ref{tab:significance}. The models with non-trivial asymmetries consistently show an increased combined significance for the 2D histograms compared with using $m_{tt}$ alone, illustrating that asymmetry observables can be used to not only aid the diagnostic capabilities provided by the cross section, but also to increase the combined significance for first discovery.

We believe that our results represent a significant phenomenological advancement in proving that charge and spin asymmetry observables can have a strong impact in accessing and profiling $Z'\to t\bar t$ signals during Run 2 of the LHC, with plans to demonstrate this further with parton-shower, hadronisation, and detector reconstruction in a forthcoming publication. This is all the more important in view of the fact that several BSM scenarios, chiefly those assigning a composite nature to the recently discovered Higgs boson, embed one or more $Z'$ state which are strongly coupled to top (anti)quarks~\cite{barducci2012}.

\begin{table}[H]
  \centering
  \begin{tabular}{|llccc|}
    \hline
    \bigstrut
    Class & U$(1)'$ & \multicolumn{3}{c|}{Significance ($Z$)} \\
    \cline{3-5}
    \bigstrut
    & & $m_{tt}$ & $m_{tt}$ \& $\cos\theta^{*}$ & $m_{tt}$ \& $\cos\theta_l$ \\
    \hline
    \bigstrut[t]
    \multirow{6}{*}{E$_6$} & U$(1)_\chi$    & $ 3.7$ & -       & -      \\
                           & U$(1)_\psi$    & $ 5.0$ & -       & -      \\
                           & U$(1)_\eta$    & $ 6.1$ & -       & -      \\
                           & U$(1)_S$       & $ 3.4$ & -       & -      \\
                           & U$(1)_I$       & $ 3.4$ & -       & -      \\
                           & U$(1)_N$       & $ 3.5$ & -       & -      \\
    \hline
    \bigstrut[t]
    \multirow{4}{*}{GLR}   & U$(1)_{R }$    & $ 7.7$ & $ 8.5$  & $ 8.6$ \\
                           & U$(1)_{B-L}$   & $ 3.6$ & -       & -      \\
                           & U$(1)_{LR}$    & $ 5.1$ & $ 5.6$  & $ 5.8$ \\
                           & U$(1)_{Y }$    & $ 6.3$ & $ 6.8$  & $ 7.0$ \\
    \hline
    \bigstrut[t]               
    \multirow{3}{*}{GSM}   & U$(1)_{T^3_L}$ & $12.1$ & $13.0$  & $14.0$ \\
                           & U$(1)_{SM}$    & $ 7.1$ & $ 7.3$  & $ 7.6$ \\
                           & U$(1)_{Q}$     & $24.8$ & -       & -      \\
    \hline
  \end{tabular}
  \caption{Expected significance, expressed as the Gaussian equivalent of the $p$-value.}
  \label{tab:significance}
\end{table}

\Acknowledgements

We acknowledge the support of ERC-CoG Horizon 2020, NPTEV-TQP2020 grant no. 648723, European Union. DM is supported by the NExT Institute and an ATLAS PhD Grant, awarded in February 2015. SM is supported in part by the NExT Institute and the STFC Consolidated Grant ST/L000296/1. FS is supported by the STFC Consolidated Grant ST/K001264/1. We would like to thank Ken Mimasu for all his prior work on $Z'$ phenomenology in $t\bar{t}$, as well as his input when creating the generation tools used for this analysis. Thanks also go to Juri Fiaschi for helping us to validate our tools in the case of Drell-Yan $Z'$ production. Additionally, we are very grateful to Glen Cowan for discussions on the statistical procedure, and Lorenzo Moneta for aiding with the implementation.

\bibliography{references}{}
\bibliographystyle{h-physrev}

\end{document}